\documentclass[lettersize,journal]{IEEEtran}
\usepackage{mathtools}
\usepackage{amssymb}
\usepackage{graphicx}
\usepackage{algorithm}
\usepackage{algpseudocode}
\usepackage{array}
\usepackage{booktabs}
\usepackage{multirow,multicol,tabularx}
\usepackage{makecell}
\usepackage{colortbl}
\usepackage{xcolor}
\usepackage{enumitem}
\usepackage{etoolbox}
\usepackage{mdframed}
\usepackage{wrapfig}
\usepackage{stfloats}
\usepackage{textcomp}
\usepackage{url}
\usepackage{cite}
\usepackage[caption=false,font=normalsize,labelfont=sf,textfont=sf]{subfig}

\begin{document}

\title{A Personalized and Adaptive Intelligent Tutoring System for Psychomotor Learning}
\author{Madeleine S. Yuh and Neera Jain
\thanks{Manuscript received September 18, 2026; revised XXX XX, 20XX. (Corresponding author: Madeleine S. Yuh)}
\thanks{Madeleine S. Yuh and Neera Jain are with the School of Mechanical Engineering, Purdue University, West Lafayette, IN, 47906, USA (e-mail: myuh@purdue.edu; neerajain@purdue.edu).}
}



\markboth{Journal of \LaTeX\ Class Files,~Vol.~14, No.~8, August~2021}%
{Shell \MakeLowercase{\textit{et al.}}: A Sample Article Using IEEEtran.cls for IEEE Journals}

\IEEEpubid{0000--0000/00\$00.00~\copyright~2026 IEEE}

\maketitle

\begin{abstract}
 Intelligent tutoring systems (ITSs) have advanced learner motivation and personalization in cognitive domains, but extending these gains to psychomotor tasks
 remains challenging, particularly for maintaining motivation and tailoring instruction to individual learners. Existing psychomotor ITSs typically rely on learner models built before instruction or updated only once per module, leaving open the question of how an ITS should use ongoing, real-time assessments of learner cognitive state to drive finer-grained personalization. More specifically, during psychomotor learning, multiple tutoring objectives should be met, including calibrating the learner's self-confidence to their performance, calibrating their workload according to the Yerkes-Dodson Law, and driving their skill progression towards higher learning stages. However, how to balance these tutoring objectives for individual learners, and how to adjust that balance as a learner progresses, remain open questions. 
 This study proposes and evaluates an online personalization algorithm that continuously reprioritizes tutoring objectives to match the evolving needs of individual learners during training. Rather than treating tutoring objectives as equally weighted throughout instruction, the proposed actor-critic-inspired algorithm dynamically updates reward weights based on recent measures of learner self-confidence, workload, and performance. These personalized weights inform a multi-input, multi-output cognitive control framework that synthesizes learner-specific tutoring policies via model-based planning. The approach was evaluated in a human-subject study ($N = 40$) using a 2D quadrotor landing task. Clustering analysis of the learned reward weights revealed distinct prioritization patterns across learners, indicating that the algorithm captures heterogeneity in learner needs and provides a principled mechanism for individualized instructional adaptation. Learners whose reward weights shifted more substantially over training tended to achieve higher final performance, and those whose weights consistently emphasized workload calibration achieved the strongest outcomes overall. These findings contribute new technical knowledge on online learner modeling and control-based personalization in psychomotor ITSs, while demonstrating how dynamically adaptive tutoring technologies can better support learning by aligning instructional decisions with learners' evolving cognitive and performance states.

\end{abstract}


\section{Introduction} \label{sec:Introduction}
Intelligent tutoring systems (ITS) aim to personalize education for learners by emulating human tutors through algorithmic frameworks that draw on models of human learning and data collected from the learner \cite{woolf_building_2008, doroudi_mastery_2020}. ITSs have been successful in closing the feedback loop by considering learner behavior in cognitive learning domains such as language acquisition, computer programming, and mathematics  \cite{mousavinasab_intelligent_2021}. In particular, many of these consider increasing learner motivation by supporting self-efficacy, a learner’s self-perception of their competence and achievements \cite{bandura_self-efficacy_1999},  as increased learner self-efficacy has been consistently associated with improved learning performance and heightened learner motivation \cite{bandura_self-efficacy_1999}. 

While ITSs have predominantly been used for cognitive learning domains, there have been several recent developments to advance ITSs designed for psychomotor tasks. A psychomotor task is defined as any motor action that is dependent on conscious mental processes \cite{laine_systematic_2022,sottilare_extending_2015, santos_training_2016}. Recent psychomotor ITS development has largely focused on structured knowledge representation and the design of adaptive, self-improving tutoring systems that enhance tutor–learner interaction. Selfit \cite{neagu_selfit_2022} enables users to self-report fatigue, motivation to train, sleep quality, and stress levels before each training session. During training, users may manually record exercise failures and report whether additional repetitions could have been performed, including the number of additional repetitions. After completing a session, users can evaluate the perceived difficulty of the training session. This information is then used by Selfit to update the training curricula. Another example is GIFT, the generalized intelligent framework for tutoring \cite{sottilare_extending_2015}, which updates a model of the student using historical learner information, self-reports, performance data, and sensor-derived behavioral and physiological measures. In addition to this, the framework leverages a motivation assessment tool (MAT) over longitudinal training.  Together, these measurements are utilized to inform GIFT on how to adapt via tutoring strategies according to the INSPIRE (Intelligent, Nurturant, Socratic, Progressive, Indirect, Reflective, Encouraging) model of tutoring success \cite{lameier_motivational_2018}. In prior work \cite{yuh_designing_2025}, an alternative approach is proposed, in which learner engagement and motivation are decomposed into the underlying cognitive states of self-confidence and workload. Seminal learning literature indicates that self-
confidence and workload directly influence self-efficacy.

Additionally, these cognitive factors are widely recognized as key determinants of human–automation interaction quality \cite{riley_operator_1996, parasuraman_automation_1996, parasuraman_humans_1997}. Therefore, to model evolution during learning, a Markov Decision Process (MDP) framework is used to probabilistically capture the dynamics of self-confidence, workload, and learning stages characterized in by the authors in prior work \cite{yuh_classification_2024}. The action space includes assistance provided during task execution and the tone of the formative feedback delivered after each task attempt. The MDP framework enables the representation of qualitative tutoring objectives through reward function design, providing a quantitative basis for synthesizing instructional policies that address multiple learning objectives.

Despite these advancements, existing psychomotor ITS are not fully realizing the goal of personalization, particularly when it comes to adaptation to achieve cognitive objectives. Supporting learner motivation in this setting requires online assessment of learner self-efficacy along with targeted interventions that can effectively influence the learner. Although researchers have recognized the need for intelligent tutoring systems (ITSs) to adapt to human cognitive behavior in cognitive learning domains \cite{erumit_design_2020,woolf_building_2008, mousavinasab_intelligent_2021, shute_intelligent_1994}, an open question remains: \textit{how should an ITS use these ongoing assessments to improve personalization strategies in psychomotor domains?} 
Currently, personalization requires an initial model of the individual student, like that provided by the aforementioned motivational assessment tool in GIFT \cite{lameier_motivational_2018} and the self-reports collected by Selfit \cite{neagu_selfit_2021}.  Both are implemented before instruction, or can also be implemented once per distinct module or task. However, the personalization of psychomotor ITSs could benefit from \textit{more frequent online updates to the model of student behavior}, especially as psychomotor tasks increase in complexity. Furthermore, the tutoring policies employed by these psychomotor ITS should be updated according to the evolving needs of a given student by leveraging adaptive control methods. This allows online updates to both the model and the policies tutoring agents use when helping the learner and would further enhance the tutoring strategies chosen by current ITSs. One example is adapting how and when information is presented by the ITS according to an individual's learning style \cite{phobun_adaptive_2010}.

The contribution of this paper is an adaptive algorithm that updates the a psychomotor ITS tutoring policy via reprioritization of tutoring objectives during the learning process for each learner. The actor-critic algorithm from reinforcement learning inspires the algorithm and considers both performance and cognitive states (self-confidence and mental workload) to determine how well the student and ITS interaction is meeting tutoring objectives. This is then used to solve for reward weights, in which the weights represent the prioritizations of each tutoring objective. The proposed algorithm leverages a learner model and multi-objective tutoring policy designed for psychomotor ITS described in the authors' prior work \cite{yuh_designing_2025, yuh_designing_2026}. The main research questions we answer are: \textbf{(RQ1)} Does the algorithm give rise to distinct patterns in the tutoring objective weights across all learners? \textbf{(RQ2)} Do the reward weights dynamically change to adapt to the learner’s needs? \textbf{(RQ3)} Is the personalization algorithm successful in prioritizing the tutoring objectives learners need? Our findings show that the personalization algorithm produces distinct patterns in the tutoring objective prioritization across learners. Through statistical analysis, the algorithm is shown to appropriately prioritize tutoring objectives according to the learners' needs.

The paper is organized in the following order. The necessary preliminaries, including the chosen psychomotor task, learning metrics, and model training, are described in Section \ref{sec:preliminaries}. We then describe policy synthesis, followed by further detail on personalization in Section \ref{sec:personalization}. In Section \ref{sec:experiment}, we describe the experiment implemented to evaluate the proposed personalization. Data processing and results are described in Section \ref{sec:dataanalysis}, followed by a discussion of the results in Section \ref{sec:discussion}. Finally, conclusions and comments for future work are in Section \ref{sec:conclusion}.

\section{Preliminaries} \label{sec:preliminaries}
In this section, we provide the necessary background for the reader. First, we describe the psychomotor task considered in this work in Section \ref{sec:task}, followed by the definition of two key learning metrics for the task in Section~\ref{sec:Learning_Metrics}. We then describe the model training approach employed to obtain a learner model capturing the dynamics of self-confidence, workload, and learning stage progression. These preliminaries are based upon existing work by the authors \cite{yuh_classification_2024, yuh_designing_2025, yuh_designing_2026}.

\subsection{Psychomotor Task: 2D quadrotor landing simulator}
\label{sec:task}
\begin{figure*}[t]
    \centering
    \subfloat[2D quadrotor landing simulator.\label{fig:experiment}]{%
        \includegraphics[width=0.48\textwidth]{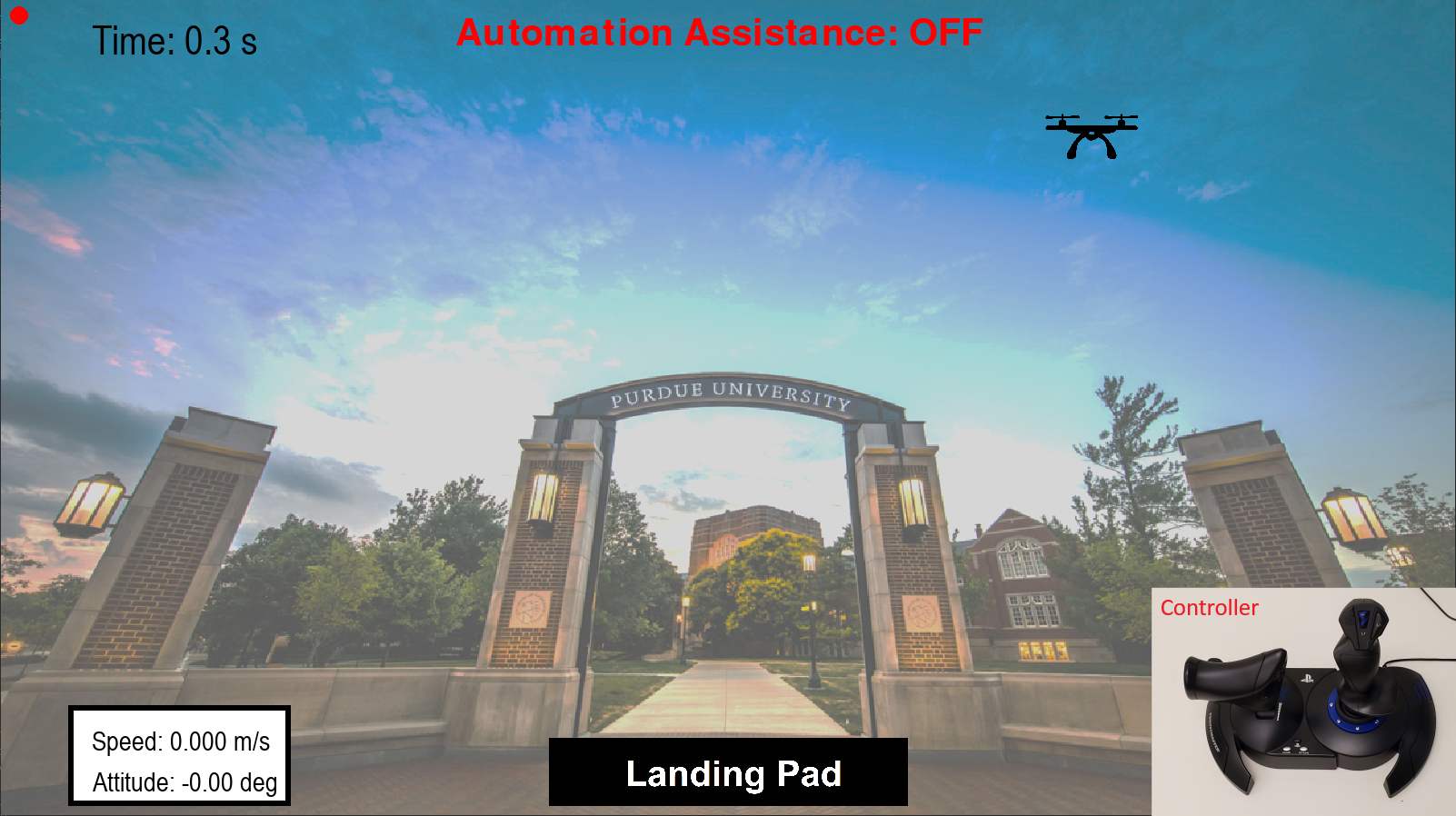}%
    }
    \hfill
    \subfloat[Quadrotor position trajectories for canonical distributions of each learning stage.\label{fig:learningstage}]{%
        \includegraphics[width=0.48\textwidth]{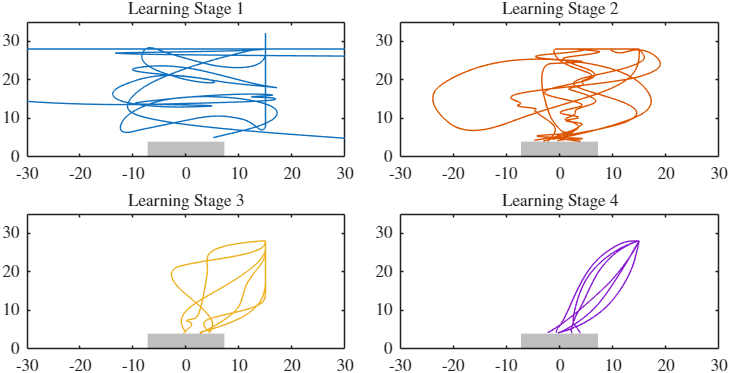}%
    }
    \caption{The designed psychomotor task.}
    \label{fig:psychomotor task}
\end{figure*}

The psychomotor task considered in this work is a 2D quadrotor landing simulator. The learner's objective is to safely guide the quadrotor onto a landing pad while satisfying constraints on its speed ($|v| \leq 15 \frac{m}{s}$) and roll angle ($|\phi|\leq 5^\circ$), using the desktop simulator shown in Fig.~\ref{fig:experiment}. The quadrotor's state consists of its position ($x$, $y$), velocity ($\dot{x}$, $\dot{y}$), roll angle $\phi$, and roll angular velocity $\dot{\phi}$. Participants control the quadrotor through a T.Flight Hotas 4 controller, using the throttle to command thrust force, $u_{1}$, and the joystick to command roll angular acceleration, $u_{2}$. A landing is classified as \emph{safe} if the quadrotor reaches the pad within the speed and roll angle limits and as \emph{unsafe} if it reaches the pad but violates one or both of these conditions. If the quadrotor crashes before reaching the pad, the attempt is considered \emph{unsuccessful}. During each trial, the participant either lands the quadrotor entirely on their own (\emph{manual} control mode, $\theta_{off}$) or receives automation assistance (\emph{shared} control mode, $\theta_{on}$). In manual mode, the participant has full control over the quadrotor's thrust force and roll angular acceleration inputs. Conversely, in shared control mode, the participant's inputs are supplemented by inputs from automation consisting of a linear-quadratic regulator that tracks a reference trajectory derived from expert demonstrations, using a linear time-invariant model of the quadrotor dynamics together with a quadratic cost function that drives all quadrotor states toward zero \cite{byeon_skill-level-based_2021}.

After each landing attempt, learners receive negative, neutral, or positive formative feedback generated using GPT-4o \cite{openai_chatgpt_2025}, following the approach of Jensen et al. \cite{jensen_automated_2024}, who showed that AI-generated formative feedback can meaningfully improve learning outcomes. The automation assistance and feedback tone are controllable inputs that the ITS can adjust before each trial of the task.

\subsection{Online Quantification of Learning}
\label{sec:Learning_Metrics}
To quantify learning in the psychomotor task online (i.e., after each trial), we employ two complementary metrics. The first is a discrete-valued metric defined in terms of psychomotor learning stages that is well-suited for a Markov decision process model.  The second is a continuous-valued measure based on a scoring function that enables gradient descent-based policy updates.  We describe each metric and its utility in the overall ITS in the subsections that follow.

\subsubsection{Discrete-valued Learning Stage Classification}
The Dreyfus model of skill acquisition learning stages was first proposed by \cite{dreyfus_five-stage_2004} as a means of characterizing learning progression in psychomotor tasks by outlining five sequential stages of learning a skill and has evolved in the decades since. Prior work \cite{yuh_classification_2024} developed a classifier for determining a participant's learning stage after completing a trial of the quadrotor landing task based on trajectory data associated with task execution. Specifically, the first four stages of the five-stage learning theory model proposed by Dreyfus \cite{dreyfus_five-stage_2004} are applied to the quadrotor landing task: novice (learning stage 1), advanced beginner (learning stage 2), intermediate (learning stage 3), and competent (learning stage 4). Features of the quadrotor dynamics as well as feasible landing types for each learning stage are summarized as follows for the reader's benefit. 
\begin{itemize}
    \item \textbf{Learning stage 1} corresponds to the \textit{novice} learner, who is still decomposing the task and developing an understanding of the overall goal. At this stage, the learner struggles to stabilize the quadrotor, and feasible landings are limited to unsuccessful or unsafe outcomes.
    \item \textbf{Learning stage 2} corresponds to the \textit{advanced beginner}, who begins developing strategies, identifying sub-tasks, and recognizing cues that indicate instability. The learner can now stabilize the quadrotor and complete landings, though the resulting trajectories remain inefficient and inconsistent. Feasible landings at this stage include unsuccessful, unsafe, and safe landings.
    \item \textbf{Learning stage 3} corresponds to the \textit{competent} learner, who refines their strategy, prioritizes among sub-tasks, and shows increasing consistency, though still under a high attention load. The learner can control the quadrotor and produces less variable trajectories, with feasible landings limited to unsafe and safe outcomes.
    \item \textbf{Learning stage 4} corresponds to the \textit{proficient} learner, whose strategy has become fixed and continues to improve with practice, accompanied by a decreasing attention load. The learner maintains control and lands the quadrotor consistently, producing efficient trajectories, with safe landings as the only feasible outcome.
\end{itemize}

Each learning stage definition is quantified using a representative distribution of trajectories associated with that stage; example trajectories for each learning stage are shown in Fig.~\ref{fig:learningstage}. These trajectories capture the full quadrotor states of position ($x$, $y$), velocity ($\dot{x}$, $\dot{y}$), roll angle $\phi$, and roll angular velocity $\dot{\phi}$ together with the thrust force input $u_{1}$ and roll angular acceleration input $u_{2}$. Each of these distributions is represented as a kernel embedding within a reproducing kernel Hilbert space, from which a learning stage classifier is built. For a detailed description of the development of the classifier, see \cite{yuh_classification_2024}. 

\subsubsection{Continuous-valued Performance Score}
A gamification approach is utilized to define a continuous-valued scoring function that is better suited for online policy adaptation than the discrete characterization of learning progression provided by the learning stage definitions. The total score $S$ is given by
\begin{equation}
\label{eqn:fullscore}
    S = \frac{S_{t,p}+S_{t,v}+S_{t,\theta}+S_{t,MMD_{LS_{4}}}}{4} \times 10.
\end{equation}
\noindent where the position score, velocity score, roll angle score, and proficiency score for each trial $t$ are given by $S_{t,p}$, $S_{t,v}$, $S_{t,\theta}$, and $S_{t,MMD_{LS_{4}}}$, respectively. The position, velocity, and roll angle scores quantify how close the landing conditions of a learner's performance are to the safe landing conditions. The proficiency score quantifies how similar one's landing attempt is to those of proficient learners (learning stage 4) based on trajectory data representative of proficient learners. The individual score components are defined as follows:
\begin{equation}
S_{t,p} = 
\begin{cases}
100& \text{, unsafe or safe landing} \\
100\left(1-\frac{\sqrt{x_{t}^{2}+y_{t}^2}}{45.2}\right)& \text{, unsuccessful}
\end{cases}
\label{eqn:Position_Score_var}
\end{equation}
\begin{equation}
S_{t,v} = 100\left(1-\frac{1}{1+e^\frac{-(v_{t}-8.5)}{2.0}}\right)
\label{eqn:Velocity_Score_var}
\end{equation}
\begin{equation}
S_{t,\theta} = 126.4\left(1-\frac{1}{1+e^\frac{-|\theta_{k}|-20.0)}{15.0}}\right) \enspace .
\label{eqn:Attitude_Score_var}
\end{equation}
\begin{equation}
S_{t,MMD_{LS_{4}}} = 
\begin{cases}
    100 &\text{, if } \frac{{MMD}_{LS_{4},t}}{1.5} > 1 \\
    \frac{{MMD}_{LS_{4},t}}{1.5}\times 100 & \text{, else}
\end{cases}
\enspace .
\label{eqn:MMD_var}
\end{equation}
Eqns.~\eqref{eqn:Position_Score_var}--\eqref{eqn:Attitude_Score_var} are defined using Sigmoid functions that are parameterized such that the total score of 750 out of 1000 points is indicative of someone who has successfully landed the quadrotor, albeit not in a manner that is representative of a canonically proficient learner. In Eqn.~\eqref{eqn:MMD_var}, $MMD$ is the maximum mean discrepancy \cite{gretton_kernel_2012} between the landing attempt trajectory and canonical distribution representing learning stage 4 ${{MMD}_{LS_{4},t}}$.

\subsection{Human Learner Model} \label{sec:model}
To design an optimal tutoring policy and adapt it online, we require a model of the human learner. Given that the tutoring objectives of interest rely on knowledge of the learner's self-confidence, workload, and learning progression, we utilize a Markov decision process (MDP) model that captures stochastic transitions between these learner states. The model is based on a previously developed and validated framework \cite{yuh_designing_2025} and is summarized here for completeness. An MDP is defined by the 5-tuple $(\mathcal{S},\mathcal{A},\mathcal{T},\mathcal{R},\gamma)$, where $\mathcal{S}$ denotes a finite set of states and $\mathcal{A}$ a finite set of actions. The transition function $\mathcal{T}$ represents the probability of transitioning to the next state $s'$ given the current state $s$ and action $a$. The MDP framework is particularly useful for designing a tutoring policy because tutoring objectives can be captured using the reward function $\mathcal{R}(s,a)$, which is explicitly a function of the learner's states.

The model captures the dynamics of cognitive states of self-confidence and workload, as well as a learning progression state defined by learning stage. Each state is treated as independent, allowing separate models that map the actions $\theta$ and $\tau$ onto each state to be trained independently. Moreover, the transition probabilities for all state combinations are given by

\begin{equation} \label{eqn:Transitions}
\begin{split}
    \mathcal{T}(s'|s,a) = \\
    \mathcal{T}(s'_{SC}|s_{SC}, a_{\theta} ,a_{\tau}) 
    \mathcal{T}(s'_{W}|s_{W}, a_{\theta} ,a_{\tau}) 
    \mathcal{T}(s'_{LS}|s_{LS}, a_{\theta} ,a_{\tau}) \enspace . \
\end{split}
\end{equation}
This independence assumption has been supported previously by \cite{yuh_designing_2025}. The self-confidence and workload states,$s_{SC}$ and $s_{W}$ respectively, are each discretized into three levels of low, medium and high.

The discretization thresholds are set at the $33\%$ and $66\%$ quantiles of the self-reported data, which is collected on a 0--100 numerical scale; each self-report is then sorted into its corresponding discrete state based on these thresholds. The learning stage MDP instead defines three states $s_{LS}$: learning stage 1 (${LS}_{1}$), learning stage 2 (${LS}_{2}$), and a combined category for learning stages 3 and 4 (${LS}_{3,4}$). Stages 3 and 4 are merged into a single category by design, due to the similarity of quadrotor trajectory behavior in both stages.

Across all models, the action set represents the decisions available to the tutoring agent. The automation action $a_\theta$ indicates whether within-trial automation assistance is off ($\theta_{off}$) or on ($\theta_{on}$), while the feedback tone action $a_{\tau}$ indicates whether the feedback delivered is negative ($\tau_{neg}$), neutral ($\tau_{neu}$), or positive ($\tau_{pos}$). 

For the reader's reference and convenience, we describe the model training method in Appendix \ref{sec:model_training}.
\section{Tutoring Policy Design and Adaptive Personalization} \label{sec:personalization}
In order to personalize the psychomotor tutoring policy to individual learners, we need a tutoring policy structured in a way that enables adaptation, as well as the adaptation algorithm itself. In this section, we first introduce the design of a weighted reward function that captures the three tutoring objectives of interest. We then propose an algorithm for adapting the weights in the reward function based on individual performance over consecutive trials.

\subsection{Reward Function Design and Policy Training}
We aim to achieve the following three tutoring objectives with a psychomotor ITS: (1) calibrating self-confidence to learning stage; (2) calibrating workload in accordance with the Yerkes-Dodson law \cite{yerkes_relation_1908}; and (3) rewarding advancement in learning stages from novice toward proficient. The first two objectives are motivated by the self-efficacy literature \cite{benabou_self-confidence_2005,hung_designing_2013, keller_development_1987, bandura_self-efficacy_1977}, which holds that greater self-efficacy makes learners more realistically self-aware about their progress and accomplishments, in turn boosting performance and motivation \cite{bandura_self-efficacy_1977,bandura_self-efficacy_1999}. Maintaining appropriately calibrated self-confidence and workload levels are thus both components of self-efficacy. The third objective explicitly encourages progression through the learning stages. Therefore utilize a multi-objective and weighted reward function $\bar{\mathcal{R}}(s,a)$ defined as
\begin{equation} \label{eqn:R(s,a)}
\begin{split}
    \bar{\mathcal{R}}(s,a) = \alpha_{1}R_{1}+ \alpha_{2}R_{2} + \alpha_{3}R_{3} \enspace ,
\end{split}
\end{equation}
where $\alpha_1$, $\alpha_2$, and $\alpha_3$ are weights reflecting the relative priority assigned to each tutoring objective, and $R_{1}$, $R_{2}$, and $R_{3}$ are the components of the total reward corresponding to three tutoring objectives. 
For brevity, detailed descriptions of $R_{1}$, $R_{2}$, and $R_{3}$ are provided in Appendix~\ref{app:reward}. The optimal policy is computed using the Bellman optimality equations \cite{sutton_reinforcement_1998,adhikari_data_2022}. The action-value function $Q(s,a)$ is given by
\begin{equation}
\begin{split}
    & Q(s,a)= \mathcal{R}(s,a)+\gamma\sum_{\substack{s'\in\mathcal{S}}}\mathcal{T}(s'|s,a)\sum_{a'\in \mathcal{A}}Q(s',a')
\end{split}
\label{eqn:Q}
\end{equation}
where $\mathcal{T}(s'|s,a)$ and $\mathcal{R}(s,a)$ correspond to Eqn.~\eqref{eqn:Transitions} and Eqn.~\eqref{eqn:R(s,a)}, respectively. The optimal policy $\pi(a|s)$ is then obtained by selecting, for each state, the action that maximizes $Q(s,a)$. As discussed in the Introduction, different learners may require more emphasis on one tutoring objective based on their individual characteristics. Moreover, even for the same learner, the appropriate prioritization will change over time. This motivates the design of a reward weight adaptation scheme, discussed next.

\subsection{Reward Weight Adaptation Algorithm}
To update the reward weights online, one may initially turn to online reinforcement learning (RL) algorithms such as actor-critic \cite{konda_actor-critic_1999}. However, in the case of the psychomotor ITS, the available dataset is not well-suited for online reinforcement learning. First, the data is sparse and collected in short, fixed-length episodes (e.g., the most recent $t$ trials), which limits the ability to perform long-horizon credit assignment or reliably estimate value functions from interaction data alone. Second, the data are generated under a restricted set of previously deployed policies, resulting in limited state--action coverage and insufficient exploration for stable policy improvement. Third, learner behavior is inherently variable and non-stationary, as performance, self-confidence, and workload evolve, violating the stationarity assumptions commonly required for RL. Finally, the human-in-the-loop nature of the system makes aggressive exploration undesirable, as frequent or suboptimal policy updates may negatively impact learning outcomes and user experience.

To address these challenges, we propose a data-driven reward re-weighting approach that leverages recent observations to adapt the relative importance of tutoring objectives, followed by model-based $Q$-iteration to compute the corresponding policy. This framework can be interpreted as an actor--critic-inspired architecture, in which the reward weights $\boldsymbol{\alpha}$ serve as high-level actor parameters that modulate the reward structure underlying policy optimization, while the objective functions (described in Sec.~\ref{sec:obj_func}) computed from recent learner data provide a critic-like signal that evaluates the extent to which the tutoring objectives are satisfied. Unlike classical actor--critic methods, however, policy updates are not performed through online interaction; instead, the policy is recomputed via model-based planning under the learned reward function, resulting in a more stable and data-efficient approach.

\subsubsection{Objective Function Definition}
\label{sec:obj_func}
To adapt the relative prioritization of tutoring objectives, we define three objective loss functions $\mathcal{V}_1,\mathcal{V}_2,\mathcal{V}_3$ that correspond directly to the reward components $R_1, R_2, R_3$. These objective functions encode the same three tutoring objectives defined in the reward function and provide a differentiable signal for adjusting the reward weights. We utilize self-reported self-confidence $\mathrm{SR}_{SC}$ and self-reported workload $\mathrm{SR}_{W}$ and scaled score $J$. Scaled score $J$ is the total score $S$ (Eqn.~\eqref{eqn:fullscore}) divided by 10, such that self-reported self-confidence,  self-reported workload, and scaled score are all in the range $[0,100]$. The objective functions are computed from recent trial history, specifically the last ten trials, using the empirical mean of self-reported self-confidence $\overline{\mathrm{SR}_{SC}}$, empirical mean of self-reported workload $\overline{\mathrm{SR}_{W}}$, and empirical mean of scaled score $\overline{J}$, together with the scaled score standard deviation $\sigma_J$.

The first objective function aligns learner self-confidence with observed performance and is defined as the squared difference of $\overline{\mathrm{SR}_{SC}}$ and $\overline{J}$,
\begin{equation}
\label{eqn:obj_SC}
\mathcal{V}_1 = (\overline{\mathrm{SR}_{SC}} - \overline{J})^2,
\end{equation}
\noindent such that both over-confidence and under-confidence are penalized. The second objective function promotes an inverse relationship between workload and performance. For example, as learners gain proficiency in executing the skill of landing the quadrotor, the cognitive workload should decrease. We take a similar approach to Eqn.~\eqref{eqn:obj_SC} by taking a squared difference. Therefore, we define $\mathcal{V}_2$ as
\begin{equation}
\label{eqn:obj_W}
\mathcal{V}_2 =
\left(
\overline{\mathrm{SR}_{W}} - (100 - \overline{J})
\right)^2 \enspace .
\end{equation}
\noindent
The third objective function aims to drive learners' performance to exceed a desired proficiency threshold while maintaining consistent task execution across trials. It is defined using a rectified linear unit (ReLU) function-based penalty \cite{brownlee_gentle_2019} for insufficient performance, combined with a variance penalty, given by 
\begin{equation}
\label{eqn:obj_LS}
\mathcal{V}_3 =
\max(0,\, \zeta - \overline{J})^2
+
\lambda \sigma_J \enspace ,
\end{equation}
where $\zeta$ denotes the desired performance threshold and $\lambda$ controls the strength of the variance penalty. This formulation encourages learners to maintain both high and consistent performance.

\subsubsection{Reward Weight Adaptation via Gradient Descent}
Rather than minimizing a weighted sum of the objective functions directly, we first convert the functions into a target priority distribution over reward components. This formulation ensures that \textit{if a given objective is not being met, its reward weight is increased}, thereby increasing the influence of under-satisfied tutoring objectives in subsequent policy optimization. Specifically, let $\tilde{\boldsymbol{\alpha}} = [\tilde{\alpha}_1,\tilde{\alpha}_2,\tilde{\alpha}_3]^\top$ denote the target reward-weight distribution induced by $\mathcal{V}_1$, $\mathcal{V}_2$, and $\mathcal{V}_3$, where
\begin{equation}
\label{eqn:target_alpha}
\tilde{\alpha}_i
=
\frac{\exp(\kappa \mathcal{V}_i)}
{\sum_{j=1}^{3} \exp(\kappa \mathcal{V}_j)},
\qquad i \in \{1,2,3\},
\end{equation}
and $\kappa > 0$ is a parameter that controls how strongly the objective least satisfied dominates the target distribution. Under this construction, larger values of $\mathcal{V}_i$ yield larger target weights $\tilde{\alpha}_i$, where $i$ denotes the tutoring objective. The learned reward weights are constrained to form a convex combination such that $\alpha_1 + \alpha_2 + \alpha_3 = 1, \enspace \alpha_i \ge 0,$ and are parameterized using a softmax transformation of unconstrained parameters given by $\boldsymbol{\alpha} = \mathrm{softmax}(\mathbf{z})$. 

To obtain interpretable reward-weight updates, we optimize $\boldsymbol{\alpha}$ so that it tracks the target distribution while remaining close to a prior reward-weight distribution $\boldsymbol{\alpha}_{prior}$. Therefore, the loss function used for optimizing the tutoring objective weights is
\begin{equation}
\label{eqn:loss}
\mathcal{L}_{\alpha}
=
D_{KL}(\boldsymbol{\alpha} \,\|\, \tilde{\boldsymbol{\alpha}})
+
\beta \, D_{KL}(\boldsymbol{\alpha} \,\|\, \boldsymbol{\alpha}_{prior}),
\end{equation}
where $D_{KL}(\cdot)$ denotes the Kullback--Leibler divergence \cite{kullback_information_1951} and $\beta \ge 0$ controls the strength of regularization toward the prior distribution. This formulation serves two purposes. First, the term $D_{KL}(\boldsymbol{\alpha} \,\|\, \tilde{\boldsymbol{\alpha}})$ drives the learned reward weights toward the priority distribution, ensuring that tutoring objectives that have not been met receive greater emphasis. Second, the prior regularization term discourages abrupt shifts in reward priorities to promote continuity across learner interactions. Reward-weight adaptation is completed via gradient descent applied to the unconstrained parameter vector $\mathbf{z}$ such that
\begin{equation}
\mathbf{z}^{(k+1)}
=
\mathbf{z}^{(k)}
-
\eta \nabla_{\mathbf{z}} \mathcal{L}_{\alpha}(\boldsymbol{\alpha}^{(k)}),
\end{equation}
where $\eta > 0$ is the learning rate. 

This procedure separates tutoring objective prioritization from policy optimization. Reward weights are first adapted using recent learner data, after which the resulting reward function $R_{\boldsymbol{\alpha}}$ is used within a model-based $Q$-iteration procedure to compute the optimal action-value function $Q^\star$ and the corresponding tutoring policy $\pi^\star$. Algorithm \ref{alg:loss_driven_weights_qiter} depicts the pseudo-code of the entire process used to optimize tutoring objective weights.

\begin{algorithm}[h]
\scriptsize
\caption{Reward weight learning followed by $Q$-iteration}
\label{alg:loss_driven_weights_qiter}
\begin{algorithmic}[1]

\Require Recent trial data $\mathcal{D}=\{(\mathrm{SR}_{SC,i},\mathrm{SR}_{W,i},J_i)\}_{i=1}^{k}$; reward components $\{R_1,R_2,R_3\}$; discount factor $\gamma$; transition model $\mathcal{T}(s'\mid s,a)$; prior weight vector $\boldsymbol{\alpha}_{prior}$; learning rate $\eta$, gradient steps $E$; performance threshold $\tau$; variance penalty $\lambda$; temperature parameter $\kappa$; KL regularization coefficient $\beta$

\Ensure Learned weights $\boldsymbol{\alpha}^\star$, weighted reward $R_{\boldsymbol{\alpha}^\star}$, optimal action-value function $Q^\star$, and policy $\pi^\star$

\vspace{0.5em}
\State \textbf{Compute empirical summaries:}
\State $\overline{\mathrm{SR}_{SC}} \gets \frac{1}{k}\sum_{i=1}^{k} \mathrm{SR}_{SC,i}$
\State $\overline{\mathrm{SR}_{W}} \gets \frac{1}{k}\sum_{i=1}^{k} \mathrm{SR}_{W,i}$
\State $\overline{J} \gets \frac{1}{k}\sum_{i=1}^{k} J_i$
\State $\sigma_J \gets \mathrm{Std}(\{J_i\}_{i=1}^{k})$

\vspace{0.5em}
\State \textbf{Compute objective functions:}
\State $\mathcal{L}_1 \gets (\overline{\mathrm{SR}_{SC}} - \overline{J})^2$
\State $\mathcal{L}_2 \gets (\overline{\mathrm{SR}_{W}} - (100 - \overline{J}))^2$
\State $\mathcal{L}_3 \gets \max(0,\tau-\overline{J})^2 + \lambda \sigma_J$

\vspace{0.5em}
\State \textbf{Construct target weight distribution:}
\State $\tilde{\alpha}_i \gets \dfrac{\exp(\kappa \mathcal{L}_i)}{\sum_{j=1}^{3} \exp(\kappa \mathcal{L}_j)}$, \quad $i \in \{1,2,3\}$

\vspace{0.5em}
\State \textbf{Initialize parameters:}
\State Initialize logits $z^{(0)} \in \mathbb{R}^3$

\For{$e = 0$ to $E-1$}
    \State $\boldsymbol{\alpha}^{(e)} \gets \mathrm{softmax}(z^{(e)})$
    \State $\mathcal{L}_{\alpha} \gets D_{KL}(\boldsymbol{\alpha}^{(e)} \| \tilde{\boldsymbol{\alpha}}) + \beta D_{KL}(\boldsymbol{\alpha}^{(e)} \| \boldsymbol{\alpha}_{prior})$
    \State $g \gets \nabla_z \mathcal{L}_{\alpha}$
    \State $z^{(e+1)} \gets z^{(e)} - \eta g$
\EndFor

\vspace{0.5em}
\State $\boldsymbol{\alpha}^\star \gets \mathrm{softmax}(z^{(E)})$

\vspace{0.5em}
\State \textbf{Construct weighted reward:}
\State $R_{\boldsymbol{\alpha}^\star}(s,a) \gets \alpha^\star_1 R_1(s,a) + \alpha^\star_2 R_2(s,a) + \alpha^\star_3 R_3(s,a)$

\vspace{0.5em}
\State \textbf{Run Q-iteration:}
\State Initialize $Q^{(0)}(s,a)$

\For{$t = 0$ to $T-1$}
    \ForAll{$(s,a)$}
        \State $Q^{(t+1)}(s,a) \gets R_{\boldsymbol{\alpha}^\star}(s,a) + \gamma \sum_{s'} \mathcal{T}(s'\mid s,a) \max_{a'} Q^{(t)}(s',a')$
    \EndFor
\EndFor

\vspace{0.5em}
\State $Q^\star \gets Q^{(T)}$
\State $\pi^\star(s) \gets \arg\max_a Q^\star(s,a)$

\State \Return $\boldsymbol{\alpha}^\star, R_{\boldsymbol{\alpha}^\star}, Q^\star, \pi^\star$

\end{algorithmic}
\end{algorithm}
\section{Human User Study} \label{sec:experiment}
With the proposed reward-weight adaptation algorithm and associated tutoring policy, we seek to answer the following research questions: \textbf{(RQ1)} Does the algorithm give rise to distinct patterns in the tutoring objective weights across all learners? \textbf{(RQ2)} Do the reward weights dynamically change to adapt to the learner’s needs? \textbf{(RQ3)} Is the personalization algorithm successful in prioritizing the tutoring objectives learners need?  To do this, we design and conduct a user study in which participants learn to land the quadrotor in the 2D simulator while receiving tutoring assistance guided by Algorithm~\ref{alg:loss_driven_weights_qiter}.

\subsection{Experiment Procedure}
\begin{figure*}[h]
    \centering
    \includegraphics[width=0.65\linewidth]{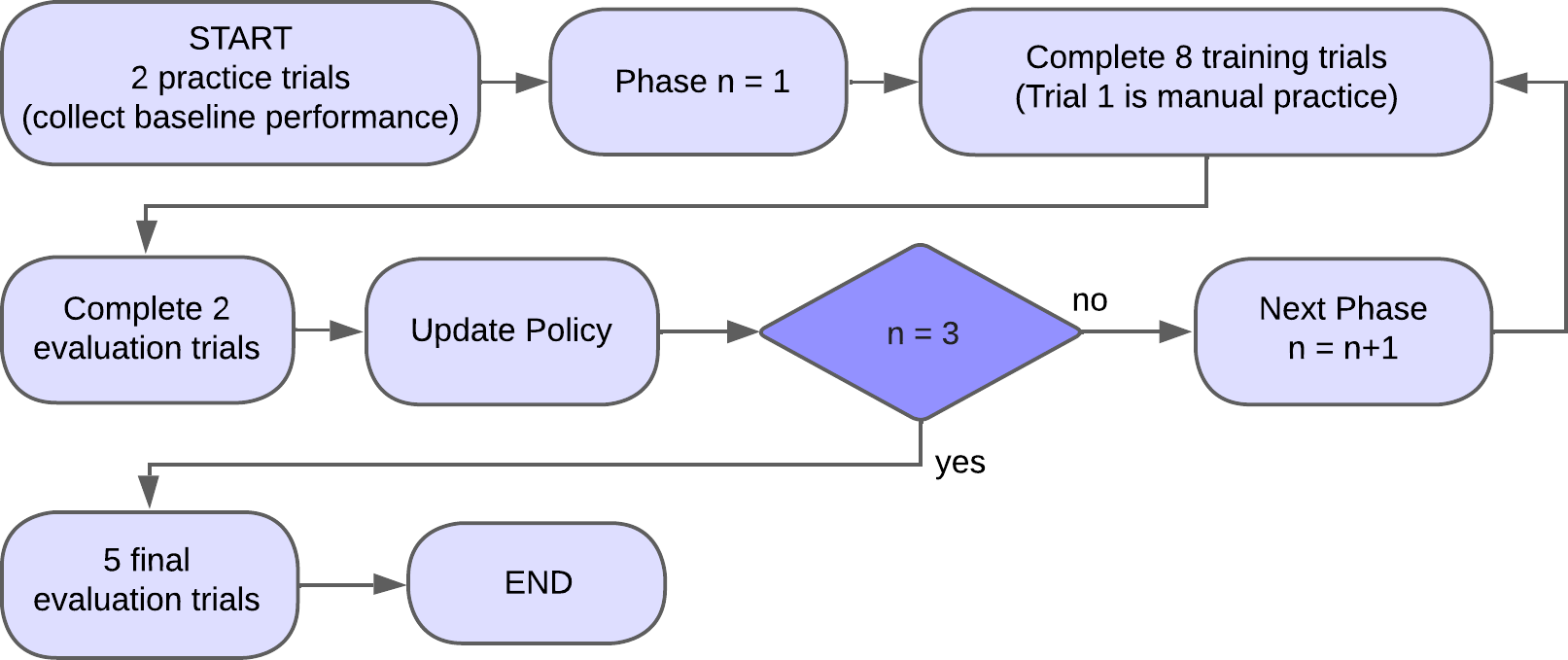}
    \caption{Experiment Flowchart}
    \label{fig:Experiment_flowchart}
\end{figure*}

Participants complete the psychomotor task described in Section~\ref{sec:task}. A flowchart depicting the sequence of events in the experiment is shown in Fig.~\ref{fig:Experiment_flowchart}. In the first two trials, participants complete the task manually and without formative feedback, providing a measurement of baseline manual performance. The next 30 trials are divided equally into three phases, where the initial position of the quadrotor changes in each phase; the sequence of the changes in position is randomized to remove ordering effects. 

At the beginning of Phase 1, the reward weights are initialized uniformly for every participant, and the associated tutoring policy is synthesized. The participant then completes one practice trial in manual mode before the tutoring policy is turned on for the next 7 trials. During these 7 trials, the policy determines whether the automation assistance should be on or off ($a_{\theta}$) at each trial as well as what tone of feedback ($a_{\tau}$) should be used when providing formative feedback to the participant. Recall that after each trial $t$, participants receive feedback on their previous landing attempt. This includes the time it took to land the quadrotor, the final speed and roll angle, and the achieved landing type. Additionally, the participants receive their total score $S$ out of 1000 points. Participants are then asked to self-report their confidence in achieving a safe landing with assistance together, as well as their perceived mental workload during the landing attempt.
For the 9th and 10th trials of Phase 1, the participant again completes two trials in manual mode. At this stage, Algorithm~\ref{alg:loss_driven_weights_qiter} is executed to update the reward weights, after which a new tutoring policy is synthesized for each participant using the Phase 1 data. For this experiment, $\zeta=85$ in Eqn.~\eqref{eqn:obj_LS}. This value is chosen based on the experimental design in which a safe landing, with a trajectory that differs substantially from expert behavior, yields at most a score of $750/1000$. Therefore, a threshold of $850/1000$ indicates that the learner’s trajectory is beginning to resemble that of a proficient operator. The scaling parameter $\kappa$ of Eqn.~\eqref{eqn:target_alpha} is set to 0.25 so that the least satisfied objective receives sufficient emphasis in the reward-weight vector. Parameter $\beta$ in Eqn.~\eqref{eqn:loss} is set to 0.01 to allow the reward weights to adapt more rapidly to each learner's needs.

Phases 2 and 3 are conducted similarly to Phase 1, with the only change being the initial position of the quadrotor on the screen and the implemented policy being based on an updated reward-weight vector. After Phase 3 ends, the participant completes 5 trials in manual mode without formative feedback to provide a measure of their performance manually landing the quadrotor. 

\subsection{Participants} 
In total, 40 participants (10 male, 29 female, 1 non-binary) completed the user study with ages ranging from 18 to 70 years ($\mu=28.8$, $\sigma=13.4$).  All participants were compensated \$20/hr. The Institutional Review Board at Purdue University, USA, approved the study. 

\subsection{Apparatus}
The quadrotor landing training simulator was developed on Python 3.12.0 using Pygame 2.6.1. The simulator was adapted for this study based on the platform described in \cite{byeon_skill-level-based_2021} and uses a Thrustmaster T. Hotas 4 with joystick and throttle to control the quadrotor. 

\subsection {Data Collection}
\label{sec:data_collection}
For each participant, the collected trial-by-trial data includes self-reported self-confidence, self-reported workload, and achieved total score. From this, we use data from the evaluation trials (the last five trials, all completed manually) to calculate the following metrics. Final performance is the mean score of the last five trials completed manually. Change in performance is the difference between the mean score of the last five trials and the mean score of the first two trials, which are all completed manually. The final self-reported self-confidence and workload are computed as the mean of the self-reports collected during these final five manual trials.

In addition to the trial-by-trial data, for each participant, the learned reward-weight vectors $\boldsymbol{\alpha}=[\alpha_1,\alpha_2,\alpha_3]$ obtained after each phase are recorded. The reward-weight vector learned after Phase 1 and implemented during Phase 2 is denoted $\boldsymbol{\alpha}^{(1)}$. Likewise, the reward-weight vectors learned after Phases 2 and 3 are denoted $\boldsymbol{\alpha}^{(2)}$ and $\boldsymbol{\alpha}^{(3)}$, respectively.
\section{Data Processing and Statistical Analysis}
\label{sec:dataanalysis}

To identify patterns in the learned reward-weight vectors determined by Algorithm~\ref{alg:loss_driven_weights_qiter} across all participants, 
we apply $k$-means clustering to the learned reward-weight vectors $\mathbf{\alpha}^{(i)}$ for $i=1,2,3$
obtained after phases 1-3.
Recall that each vector represents the relative prioritization of the reward components, and each participant contributes three reward-weight vectors, one learned after each phase of the experiment.
Changes in cluster assignment between consecutive phases are interpreted as shifts in the underlying tutoring strategy. 
The number of clusters $k$ is determined using silhouette analysis \cite{rousseeuw_silhouettes_1987}, with the elbow method \cite{GeeksforGeeks_elbow_nodate} used for secondary verification. The silhouette method evaluates clustering quality by comparing, for each data point, its average distance to points within its assigned cluster against its distance to points in the nearest neighboring cluster. The resulting silhouette score lies in the range $[-1, 1]$, where higher values indicate well-separated and cohesive clusters. 

\begin{figure*}[t]
    \centering
    \subfloat[According to the silhouette tradeoff method, $k=3$ has the highest silhouette value while minimizing the number of clusters.\label{fig:kmeans_silhouette}]{%
        \includegraphics[width=0.48\textwidth]{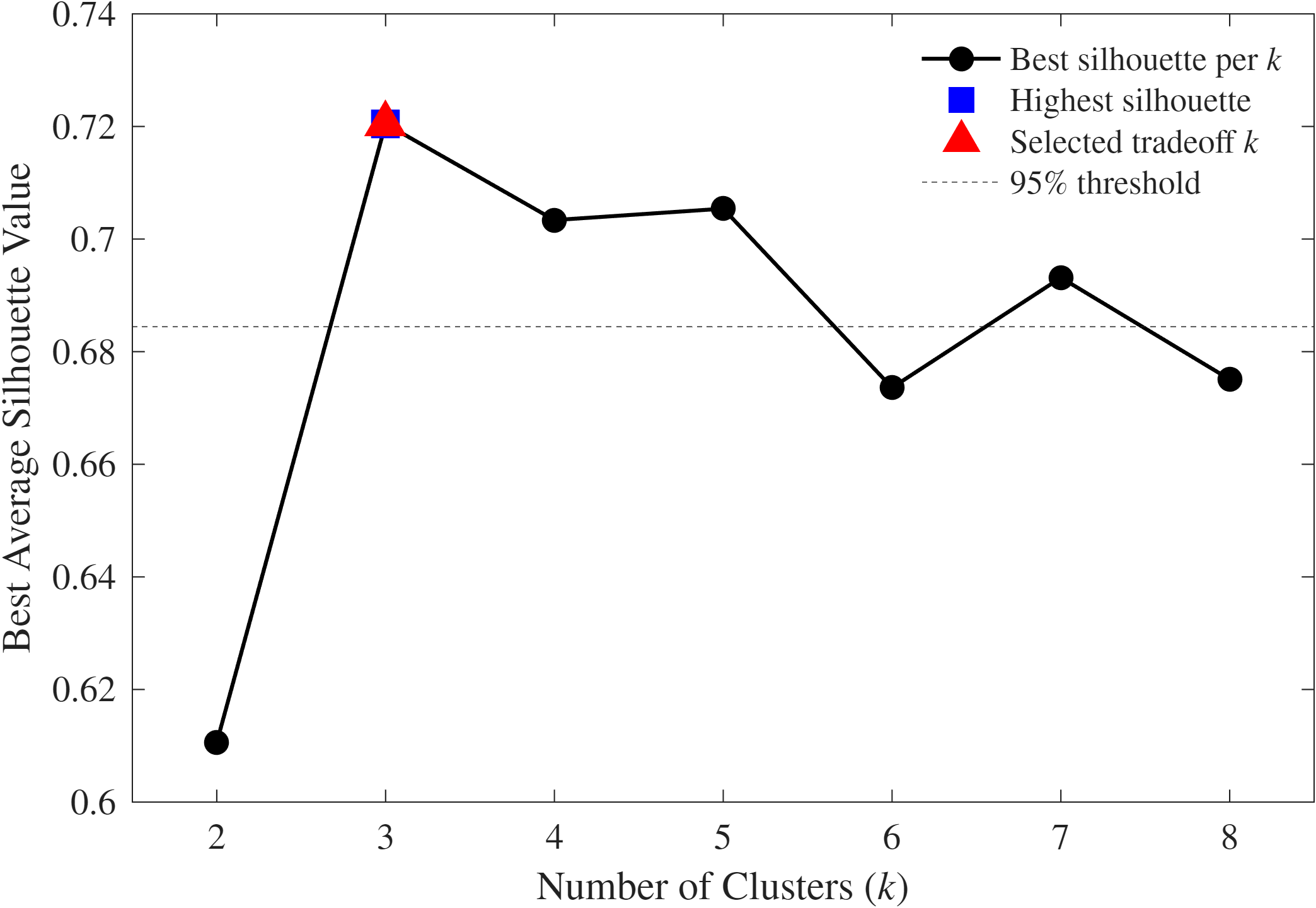}%
    }
    \hfill
    \subfloat[According to the elbow method, $k=4$.\label{fig:kmeans_elbow}]{%
        \includegraphics[width=0.48\textwidth]{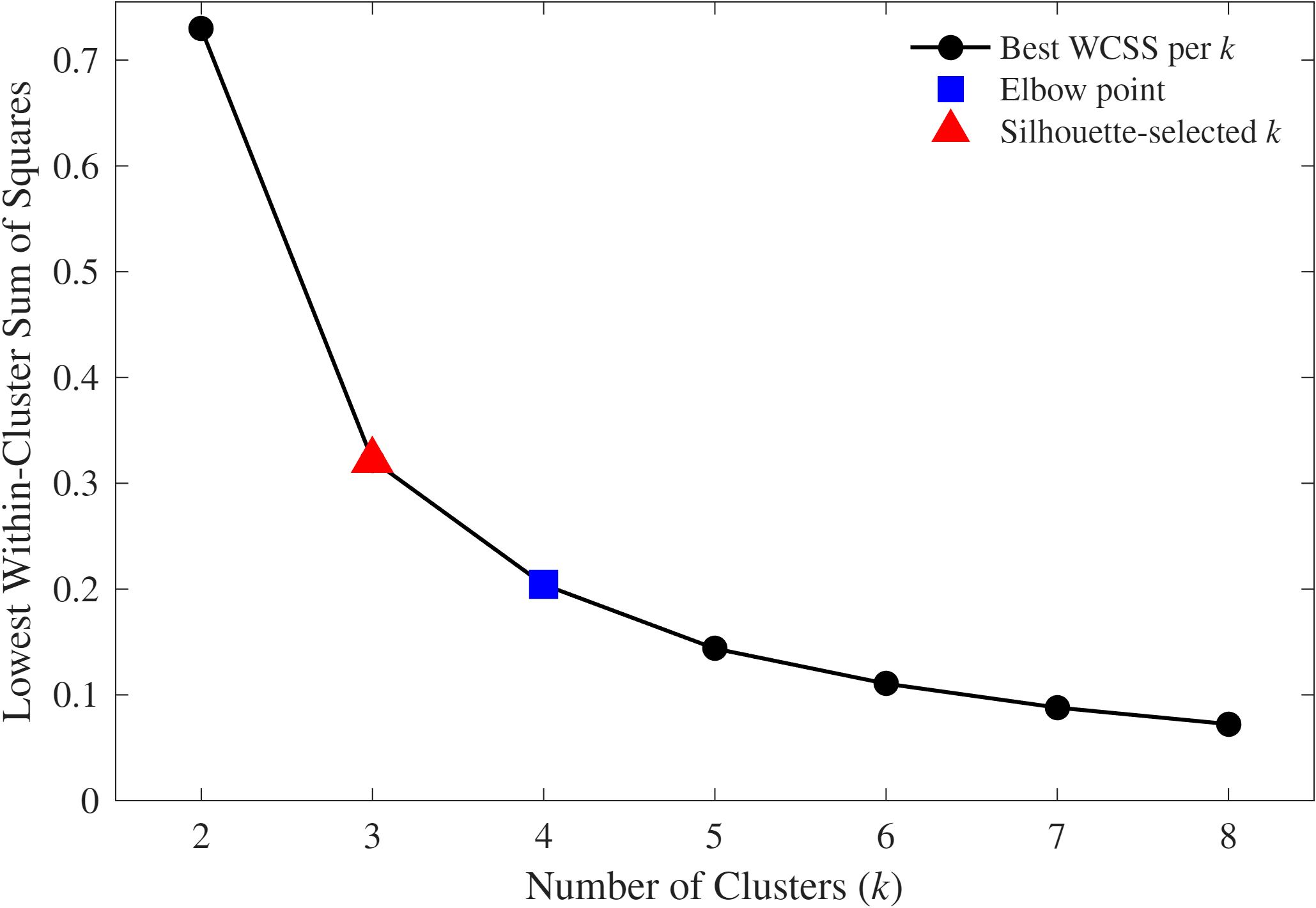}%
    }
    \caption{$k$-means clustering verification.}
    \label{fig:kmeans-verify}
\end{figure*}

\begin{figure*}[t]
    \centering
    \subfloat[Resulting $k=3$ clusters.\label{fig:k3}]{%
        \includegraphics[width=0.48\textwidth]{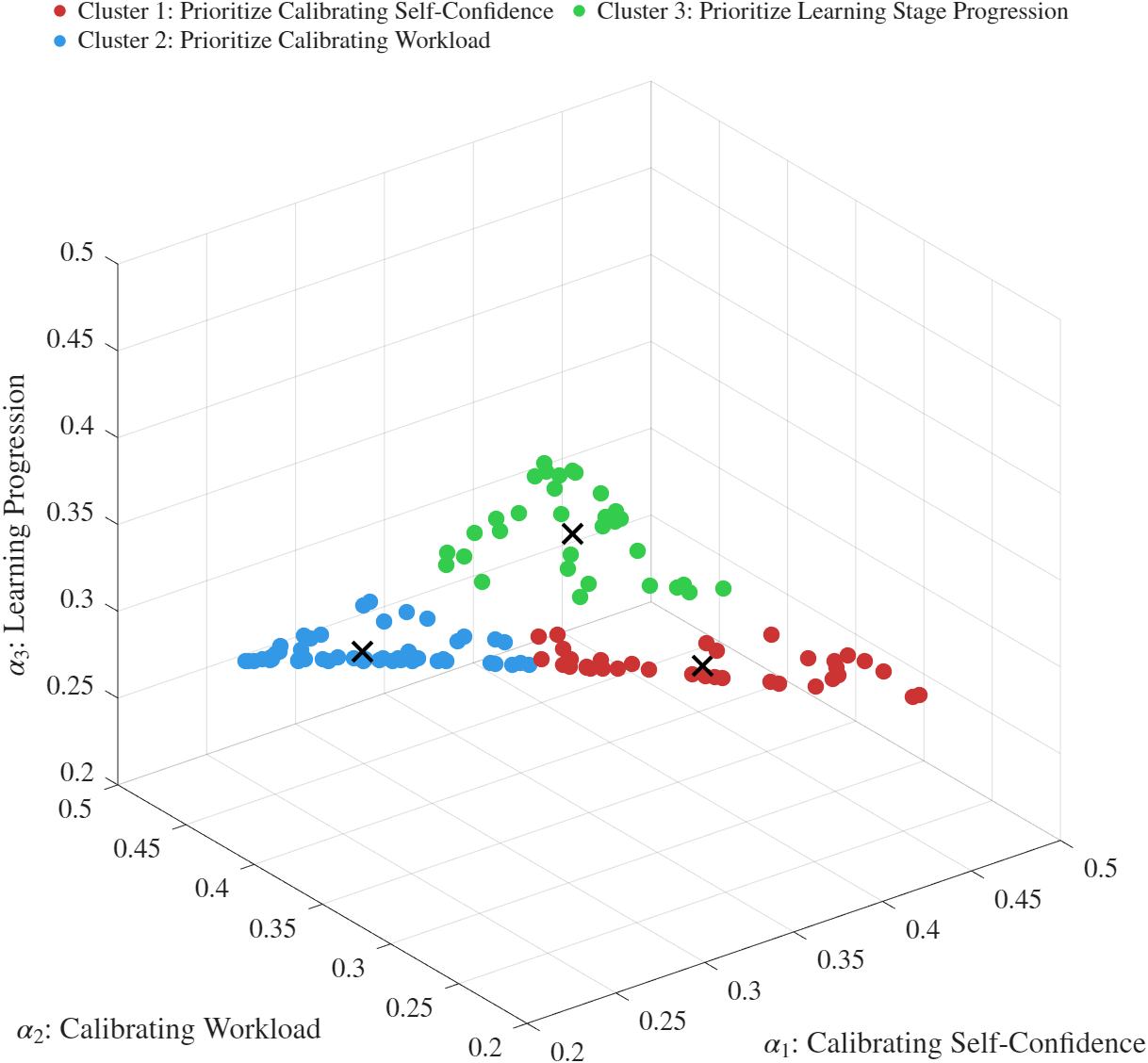}%
    }
    \hfill
    \subfloat[Resulting $k=4$ clusters.\label{fig:k4}]{%
        \includegraphics[width=0.48\textwidth]{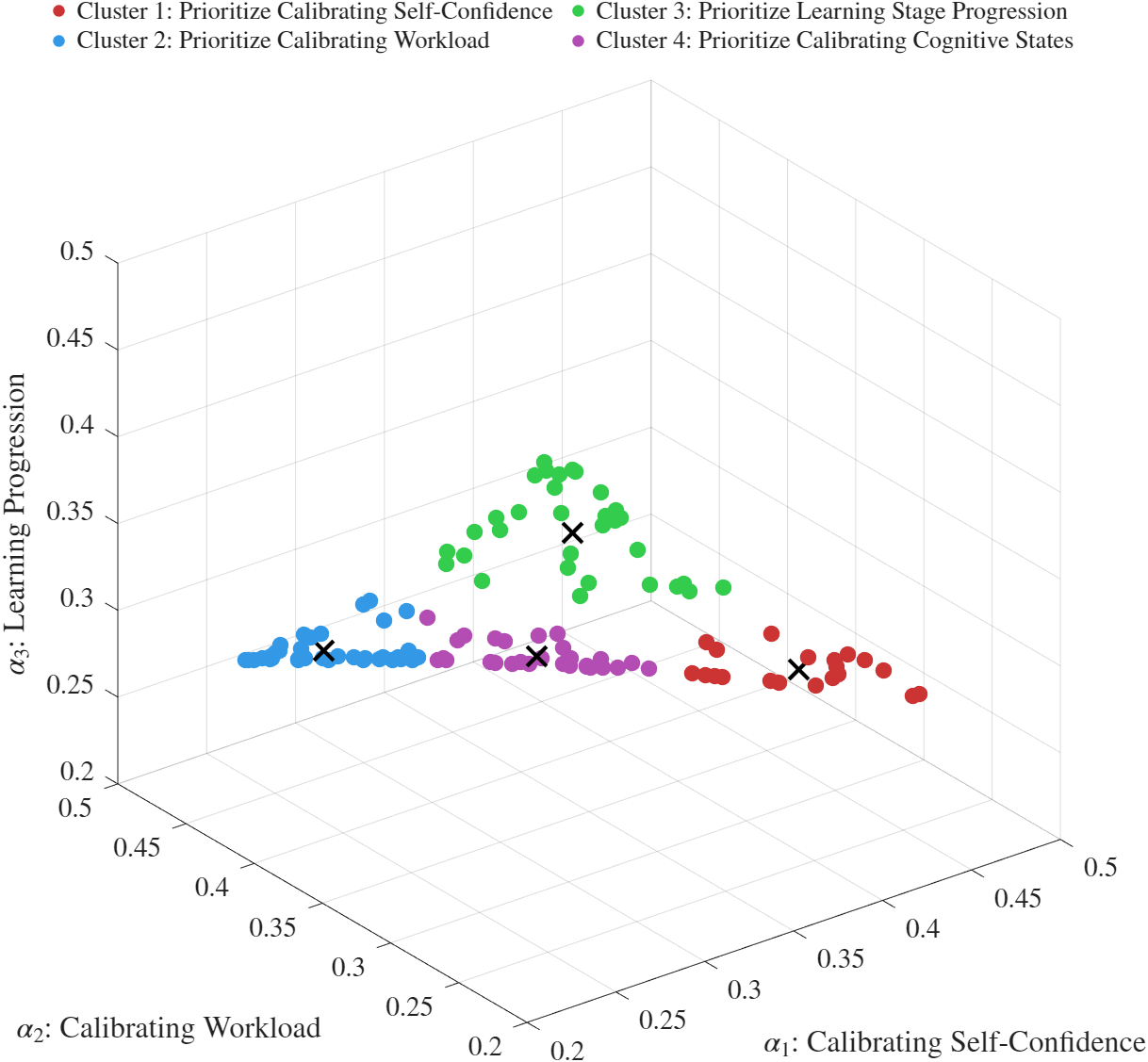}%
    }
    \caption{$k$-means clustering results for the learned reward-weight vectors. Each reward-weight vector is defined by $(\alpha_1,\alpha_2,\alpha_3)$, where $\alpha_1$ corresponds to the tutoring objective of self-confidence calibration, $\alpha_2$ corresponds to workload calibration, and $\alpha_3$ corresponds to progression through the learning stages. (a) Three-cluster solution selected using silhouette analysis. (b) Four-cluster solution obtained using the elbow method. Cluster 1 prioritizes self-confidence calibration, Cluster 2 prioritizes workload calibration, Cluster 3 prioritizes learning-stage progression, and the additional Cluster 4 in the four-cluster solution represents a balanced emphasis on self-confidence and workload calibration. The three-cluster solution was selected for subsequent analysis because it provides the best trade-off between cluster quality and model complexity.}
    \label{fig:kmeans-results}
\end{figure*}

The optimal $k$ is selected based on the average silhouette score, with preference given to the smallest $k$ that achieves near-maximum performance (within the 95\% threshold of the maximum performance) to balance clustering quality and model simplicity. The elbow method examines the within-cluster sum of squares (WCSS), defined as $\sum_{j=1}^{k} \sum_{\mathbf{x}_i \in \mathcal{C}_j} \|\mathbf{x}_i - \boldsymbol{\mu}_j\|^2$, as a function of $k$, and identifies the point at which additional clusters yield diminishing reductions in variance. This ``elbow'' provides a structural indication of an appropriate number of clusters and is used to verify the consistency of the silhouette-based selection. The results are shown in Fig. \ref{fig:kmeans-verify}. Although the elbow method suggests $k=4$ clusters and the silhouette method suggests $k=3$ clusters, the discrepancy is limited to a single cluster, indicating that both methods capture a similar underlying structure. The resulting clusters are depicted in Fig. \ref{fig:kmeans-results}.

We prioritize the silhouette-based selection, using $k=3$ clusters, as it directly quantifies clustering quality through intra-cluster cohesion and inter-cluster separation (as opposed to the elbow method, which provides a heuristic assessment based on variance reduction and may yield less distinct transition points). Cluster 1 represents prioritizing tutoring objective 1 of self-confidence calibration, cluster 2 represents prioritizing tutoring objective 2 of workload calibration, and cluster 3 represents prioritizing tutoring objective 3 of progressing in learning stage. 

We leverage the resulting cluster centroids to determine the patterns of reward-weight vectors and categorize participants into one of four groups. From the 40 participants, nine participants have consistent reward-weight vectors that prioritize tutoring objective 1 (calibrating self-confidence). Eight participants have consistent reward-weight vectors that prioritize tutoring objective 2 (calibrating workload), and seven participants have consistent reward-weight vectors that prioritize tutoring objective 3 (rewarding forward progression in learning).  Sixteen participants have reward-weight vectors that change such that the cluster to which those weights belong is different, indicating a fundamental shift in tutoring objective priority. For these participants, Table~\ref{tab:changed_phase12_summary} in Appendix~\ref{app:change_group} lists the clusters that the reward-weight vectors $\alpha^{(1)}$ and $\alpha^{(2)}$ were classified into, as well as their mean score,  self-reported self-confidence, and self-reported workload in the five final evaluation trials completed manually. For brevity, we will refer to these different groups as the SC group, W group, LS group, and change group, respectively. 

Using these cluster-based group definitions, we perform one-way ANOVA to determine whether final performance, change in performance, final self-reported self-confidence, and final self-reported workload differ across groups. The one-way ANOVA shown in Table~\ref{tab:anova_summary} allows us to investigate whether there are statistically significant differences in metrics between groups of participants. This is followed by post hoc multiple-comparison tests to determine which groups differ from one another. 
\begin{table}[h]
\centering
\caption{Summary of one-way ANOVA results across outcome variables.}
\label{tab:anova_summary}
\begin{tabular}{lcccc}
\toprule
Variable & $df_1$ & $df_2$ & $F$-value& $p$-value \\
\midrule
Final Performance & 3 & 36 & 4.1286 & \textbf{0.0129} \\
Change in Performance & 3 & 36 & 0.9578 & 0.4231 \\
Final Self-Reported Self-Confidence & 3 & 36 & 6.0819 & \textbf{0.0019} \\
Final Self-Reported Workload & 3 & 36 & 0.2877 & 0.8339 \\
\bottomrule
\end{tabular}
\end{table}
As shown in Table~\ref{tab:anova_summary}, there are significant differences in groups for the metrics of final performance and self-confidence. Therefore, post hoc multiple-comparison tests are conducted for both metrics. Tables~\ref{tab:multcompare_score} and \ref{tab:multcompare_sc} show the results of the multiple comparison test for final performance and final self-confidence, respectively. Each table lists the groups, their means $\mu$, standard deviations $\sigma$, the mean difference between groups, the lower and upper confidence intervals (CI) of the differences between groups, and the significance of the comparison via $p$-value. 
\begin{table*}[h]
\centering
\caption{Pairwise comparisons for final performance score. Score is defined in the range of 0 to 1000, and a score of 750 is required for a safe landing. A score of 850 is indicative that the learner’s trajectory is beginning to resemble that of a proficient operator.}
\label{tab:multcompare_score}
\begin{tabular}{llcccccccc}
\toprule
Group 1 & Group 2 & $\mu_{1}$ & $\sigma_{1}$ & $\mu_{2}$ & $\sigma_{2}$ & Mean diff. & Lower CI & Upper CI & $p$-value \\
\midrule
Change & SC & 814.39 & 170.96 & 793.27 & 107.95 & 21.121 & -153.605 & 195.847 & 0.9879 \\
Change & W  & 814.39 & 170.96 & 933.95 & 35.24 & -119.562 & -301.143 & 62.018 & 0.3025 \\
Change & LS & 814.39 & 170.96 & 651.66 & 235.37 & 162.730 & -27.301 & 352.761 & 0.1155 \\
SC & W & 793.27 & 107.95 & 933.95 & 35.24 & -140.683 & -344.447 & 63.081 & 0.2633 \\
SC & LS & 793.27 & 107.95 & 651.66 & 235.37 & 141.610 & -69.719 & 352.939 & 0.2878 \\
W & LS & 933.95 & 35.24 & 651.66 & 235.37 & \textbf{282.293} & 65.262 & 499.323 & \textbf{0.0065} \\
\bottomrule
\end{tabular}
\end{table*}

\begin{table*}[h]
\centering
\caption{Pairwise comparisons for self-confidence. Self-confidence is reported on a scale from 0 to 100.}
\label{tab:multcompare_sc}
\begin{tabular}{llcccccccc}
\toprule
Group 1 & Group 2 & $\mu_{1}$ & $\sigma_{1}$ & $\mu_{2}$ & $\sigma_{2}$ & Mean diff. & Lower CI & Upper CI & $p$-value \\
\midrule
Change & SC & 69.56 & 21.22 & 46.67 & 26.94 & 22.896 & -1.207 & 46.998 & 0.0677 \\
Change & W & 69.56 & 21.22 & 88.50 & 13.42 & -18.938 & -43.985 & 6.110 & 0.1939 \\
Change & LS & 69.56 & 21.22 & 55.29 & 21.55 & 14.277 & -11.937 & 40.490 & 0.4675 \\
SC & W & 46.67 & 26.94 & 88.50 & 13.42 & \textbf{-41.833} & -69.941 & -13.725 & \textbf{0.0016} \\
SC & LS & 46.67 & 26.94 & 55.29 & 21.55 & -8.619 & -37.771 & 20.533 & 0.8556 \\
W & LS & 88.50 & 13.42 & 55.29 & 21.55 & \textbf{33.214} & 3.276 & 63.152 & \textbf{0.0248} \\
\bottomrule
\end{tabular}
\end{table*}

For the reader's convenience, we summarize the means and standard deviations of final performance, final self-reported self-confidence, and final self-reported workload in Table~\ref{tab:meanandstdev}.

\begin{table*}[h]
    \centering
    \caption{Mean and standard deviations of final performance, self-reported self-confidence, and self-reported workload.}
    \begin{tabular}{lcccccc}
    \toprule
    Group & \makecell{Mean Final\\Performance} & \makecell{Final Performance\\Standard Deviation } & \makecell{Mean of\\Final ${SR}_{SC}$} & \makecell{Final ${SR}_{SC}$\\Standard Deviation} & \makecell{Mean of \\Final ${SR}_{W}$} & \makecell{Final ${SR}_{W}$\\ Standard Deviation} \\
    \midrule
        Change & 814.39 & 170.96 & 69.56 & 21.22 & 54.75 & 25.69 \\
        SC     & 793.27 & 107.95 & 46.67 & 26.94 & 52.00 & 24.92 \\
        W      & 933.95 &  35.24 & 88.50 & 13.42 & 45.75 & 20.70 \\
        LS     & 651.66 & 235.35 & 55.29 & 21.55 & 53.14 &  9.91 \\ \bottomrule
    \end{tabular}
    \label{tab:meanandstdev}
\end{table*}
\section{Results and Discussion} \label{sec:discussion}
Here we answer and discuss the three research questions posed in this work.

\subsection{RQ1: Does the algorithm give rise to distinct patterns in the tutoring objective weights across all learners?}
\textbf{Yes, distinct patterns in the tutoring objective weights across all learners are evident} based on the clustering analysis presented in Section~\ref{sec:dataanalysis}.
We first consider the case of three distinct clusters (Fig.~\ref{fig:k3}).  Cluster 1 represents prioritizing tutoring objective 1 of self-confidence calibration, cluster 2 represents prioritizing tutoring objective 2 of workload calibration, and cluster 3 corresponds to prioritizing tutoring objective 3, namely progression through the learning stages. In the case of four distinct clusters (Fig.~\ref{fig:k4}), we observe the same reward-weight vector patterns as those resulting from $k=3$, with the addition of a fourth cluster which seems to more equally weight tutoring objectives 1 and 2 of self-confidence and workload calibration, respectively. In other words, this fourth cluster prioritizes cognitive state calibration more generally, relative to helping the learner progress in the learning stage.

\subsection{RQ2: Do the reward weights dynamically change to adapt to the learner’s needs?}
\textbf{Yes. For some participants, the reward-weight vector changes after Phase 1, resulting in a dynamically updated tutoring policy that adapts to the learner's needs.} A participant is considered to exhibit a change in reward weighting if the reward-weight vectors learned for two consecutive phases are assigned to different clusters, as shown in Table \ref{tab:changed_phase12_summary} in Appendix \ref{app:change_group}. Most participants in the change group achieve a mean final evaluation score above 750/1000 among the last five trials. This is indicative of safe landings according to the score design in Eqn.~\eqref{eqn:fullscore}. The mean of the change group's mean final evaluation scores among the last five trials is 814.39 with a standard deviation of 170.96 (Table~\ref{tab:multcompare_score}). Note that given the between-subjects design of the user study, we cannot explicitly determine whether a participant in the change group benefited from the updated tutoring policy relative to experiencing a static one. However, the fact that participants in the change group did not have statistically different learning outcomes compared to the other groups (see Tables~\ref{tab:multcompare_score} and \ref{tab:multcompare_sc}) suggests that these participants benefited from the personalization.

\subsection{RQ3: Is the proposed algorithm successful in prioritizing the tutoring objectives learners need?}
\textbf{The results suggest that the personalization algorithm successfully prioritizes tutoring objectives according to learner needs.} This conclusion is supported by the statistical analysis described in Section~\ref{sec:dataanalysis}. First, we found no statistically significant difference among the groups for change in performance or final self-reported workload (Table~\ref{tab:anova_summary}). Second, as shown in Tables~\ref{tab:multcompare_score} and \ref{tab:multcompare_sc}, pairwise comparisons between groups showed that with few exceptions, there is no statistically significant differences in final performance score or final self-reported self-confidence. This suggests that each learner benefited from the personalized tutoring policy they experienced.

To further examine whether the proposed algorithm adapted the tutoring objectives correctly to each learner, we consider the analysis summarized in Table~\ref{tab:meanandstdev}.
Notably, the learning stage (LS) group exhibited the lowest mean final performance and the highest standard deviation of final performance among all groups, as shown in Table~\ref{tab:meanandstdev}. The mean final score of 651.66 is near the midpoint of the scoring scale (500/1000), and so the LS group's mean final self-confidence (55.29) and mean workload (52.00) are calibrated to performance according to the designed objective functions described in Eqns.~\eqref{eqn:obj_SC} and ~\eqref{eqn:obj_W}. \textbf{This is consistent with the interpretation that the LS group received reward weights prioritizing forward progression in learning, because the learning stage tutoring objective was the least met of the three objectives. }

In contrast, the learners whose reward-weight vectors were biased toward workload calibration achieved the highest mean final scores and the lowest standard deviation across all groups. The W group achieved a mean final score of 933.95/1000 with a standard deviation of 35.24, indicating that learners in the W group were most consistently able to achieve safe quadrotor landing trajectories consistent with learning stages 3 and 4, according to the score design in Eqn.~\eqref{eqn:fullscore}. In fact, as shown in Table~\ref{tab:multcompare_score}, the significant pairwise comparison is between the W and LS groups, with the W group achieving a significantly higher mean final score than the LS group (the lowest and least consistent performing group). Consistent with achieving the highest performance and performance consistency in the evaluation trials, the W group also reported the highest mean and lowest standard deviation of final self-reported self-confidence across all groups, as shown in Table~\ref{tab:meanandstdev}. According to Table~\ref{tab:multcompare_sc}, the W group achieved significantly higher final self-reported self-confidence than both the self-confidence (SC) and LS groups. However, the objective defined in Eqn.~\eqref{eqn:obj_W} targets low workload as performance increases; the W group reported a mean final workload of 45.75/100 with a standard deviation of 20.70, which is elevated relative to their high performance. \textbf{This miscalibration between workload and performance is consistent with the interpretation that the W group received a reward-weight vector prioritizing workload calibration throughout the experiment.}

Finally, despite achieving the second-highest mean final performance of 793.27/1000 with a standard deviation of 107.95, indicative of unsafe and safe landings categorized in learning stages 2 and 3, the SC group reported substantially lower self-confidence than expected under calibrated conditions. According to Eqn.~\eqref{eqn:obj_SC}, a calibrated self-confidence for this level of performance would be approximately 80.00/100; however, the SC group self-reported a mean self-confidence of only 46.67/100 with a standard deviation of 26.94, the highest standard deviation of self-reported self-confidence across all groups. For comparison, the W group, which achieved the highest mean final performance, reported a mean self-confidence of 88.50/100, closer to the value expected for the SC group given its performance level. As shown in Table~\ref{tab:multcompare_sc}, the SC group self-reported significantly lower mean final self-confidence than the W group, further confirming this miscalibration. \textbf{Because self-confidence was the least calibrated tutoring objective for this group, the SC group received a reward-weight vector prioritizing self-confidence calibration throughout the experiment.}

\section{Conclusion}
\label{sec:conclusion}
In this paper, we presented and evaluated an algorithm for personalizing tutoring assistance to individual learners by dynamically updating the relative prioritization of tutoring objectives within a multi-input, multi-output cognitive control framework for psychomotor intelligent tutoring systems (ITSs). The three tutoring objectives considered are aligning self-confidence with performance, calibrating workload to performance, and promoting progression through psychomotor learning stages. We enable the tutoring agent to adapt to changes in a learner’s cognitive states and learning stage via an actor–critic-inspired architecture in which the reward weights serve as high-level actor parameters that shape policy optimization, while tutoring objective-specific functions computed from recent learner data provide a critic-like signal. Unlike classical actor–critic methods, the policy is not updated through online interaction but instead recomputed via model-based planning under the learned reward function. Moreover, the algorithm overcomes limitations posed by the data requirements of traditional actor-critic reinforcement learning by using gradient descent to optimize the reward weight parameters. 

To evaluate the proposed adaptive tutoring algorithm, we designed and conducted a human user study. We showed, through $k$-means clustering, that the personalization algorithm did result in distinct patterns in the tutoring objective weights across all learners. Using these clusters, we identified which learners' reward-weight vectors changed sufficiently to move them to a different cluster after each phase of the study.  Participants whose learned reward-weight vectors changed sufficiently to transition between clusters generally achieved higher final performance than participants whose reward-weight vectors remained within the same cluster, with the exception of the workload-calibration group. 

This was followed by a statistical analysis, which showed that the personalization algorithm successfully prioritized tutoring objectives according to the observed needs of individual learners. A limitation of the results is the number of participants; despite collecting data from a total of 40 participants, we could not control how many people were categorized into each cluster (and hence the groups defined based on these clusters). As a result, some groups contained relatively few participants, with the smallest groups containing only seven participants.

Nevertheless, this work provides a computational framework and associated algorithms for achieving personalization during psychomotor learning and demonstrates, quantitatively, the benefits of this personalization. Possible directions for future work include a more extensive user study to further validate the efficacy of the proposed ITS, consideration of inter-dependencies between cognitive states and their influence on tutoring objectives, and personalization of the timescale on which reward weights are updated for individual learners.

\section*{Acknowledgments}
This material is based upon work supported by the National Science Foundation under Award No. 1836952. Any opinions, findings, and conclusions or recommendations expressed in this material are those of the authors and do not necessarily reflect the views of the National Science Foundation.

\bibliography{references}
\bibliographystyle{IEEEtran}

\appendix
\section{Model Training}
\label{sec:model_training}
Parameters for these self-confidence, workload, and learning stage MDP models were estimated by iteratively optimizing the initial and transition probabilities that maximize the likelihood of the observed state and action sequences. Optimization was performed with the genetic algorithm available in MATLAB's Optimization Toolbox \cite{noauthor_matlab_2026}.

Both states and actions are recorded at each trial, and the likelihood was computed from the joint probability of the full state-action history through trial $k$, $\mathbb{P}(s_{0:k},a_{1:k})$. Applying the chain rule together with the Markov property \cite{sutton_reinforcement_1998,adhikari_data_2022} reduces this joint probability to
\begin{equation}
    \begin{split}
        \mathbb{P}(s_{0:k},a_{1:k}) = \mathbb{P}(s_{0})\times\prod_{i=1}^{k}\mathbb{P}(s_{i}|a_{i},s_{i-1}).\\
    \end{split}
\label{eqn:MDPlikelihood}
\end{equation}
This expression gives the likelihood of the observed trajectory up through the final trial $K$, using the transition probabilities defined in Eqn. \eqref{eqn:Transitions}.

A 5-fold cross-validation is implemented to train the model. The dataset was split into five randomly assigned folds, with four folds used for training and the remaining fold for testing. The procedure is repeated so that each fold serves as the test set once. The selected model achieved the highest combined log-likelihood across both the training and test partitions.

For each model, the transition probabilities form a $|\mathcal{S}|\times|\mathcal{S}|\times|\mathcal{A}|$ array. With three states and six actions, each MDP model yields a $3\times3\times6$ array encoding all 54 possible state transitions. As an illustration, Table~\ref{tab:T-Probs} lists the transition probabilities $\mathcal{T}({SC}'|{SC},{\theta}_{on}, {\tau}_{pos})$.

\begin{table}[h]
    \centering
    \caption{Transitions probability $\mathcal{T}({SC}'|{SC},{a}_{\theta},{a}_{\tau})$ where feedback tone $\tau$ is positive and assistance $\theta$ is both off (left) and on (right).}
    \begin{tabular}{c|ccc|ccc} \toprule
         \multicolumn{1}{c|}{Actions}& \multicolumn{3}{c|}{${\theta}_{off}, {\tau}_{pos}$}&  \multicolumn{3}{c}{${\theta}_{on}, {\tau}_{pos}$} \\
         \multicolumn{1}{c|}{States} & ${SC}'_{L}$ & ${SC}'_{M}$ & ${SC}'_{H}$  & ${SC}'_{L}$ & ${SC}'_{M}$ & ${SC}'_{H}$ \\ \midrule
         ${SC}_{L}$  &  0.705  & 0.243 & 0.0528 & 0.781  & 0.153 & 0.0660\\
         ${SC}_{M}$  &  0.207 & 0.632  & 0.161  & 0.176  & 0.725 & 0.0989 \\
         ${SC}_{H}$  &  0.121 & 0.110  & 0.769  & 0.115  & 0.244 & 0.642\\ \bottomrule
    \end{tabular}
    \label{tab:T-Probs}
\end{table}

To confirm that the trained models were suitable for policy synthesis, their predictive accuracy was evaluated using multi-class receiver operating characteristic (ROC) analysis, comparing the predicted self-confidence, workload, and learning-stage classifications against participants' self-reported cognitive states and observed learning stages. Additional details of the validation procedure are available in \cite{yuh_designing_2025}.

\section{Reward Design}
\label{app:reward}
Each tutoring objective is associated with a reward function grounded in established learning-science literature. The reward functions are adapted from the formulation presented in \cite{yuh_designing_2025} and are reproduced here for completeness.

\emph{Objective 1:} The reward $R_{1}(s,a)$ combines two equally weighted components: a calibration term based on the automation-assistance action, $R(s_{SC},s_{LS},a_{\theta})$, and a calibration term based on the feedback-tone action, $R(s_{SC},s_{LS},a_{\tau})$ (Table~\ref{tab:R_SC_components}). Both terms push the learner's self-reported confidence state toward agreement with their actual learning stage. Consider $R(s_{SC},s_{LS},a_{\tau})$ as an example: a learner who reports high confidence while still at learning stage 1, a case of severe overconfidence, receives a reward of 0.60 for negatively toned feedback versus only 0.20 for positively toned feedback, reflecting that corrective tone is more appropriate in this case. Similarly, leaving automation off yields a higher reward than turning it on, such that the learner is shown that they perform worse without assistance. Table~\ref{tab:R_1} gives the resulting combined values for $R_{1}$.

\begin{table}[h]
    \centering
    \caption{Component reward functions $R(s_{SC},s_{LS},a_{\theta})$ and $R(s_{SC},s_{LS},a_{\tau})$ used to calibrate self-confidence to learning stage. Green highlighting indicates calibrated cases, while yellow and red highlighting indicate light and severe mis-calibration, respectively.}
    \label{tab:R_SC_components}
    \begin{tabular}{cc|cc|ccc}
    \toprule
    \multicolumn{2}{c}{}& \multicolumn{2}{c}{$R(s_{SC},s_{LS},a_{\theta})$} & \multicolumn{3}{c}{$R(s_{SC},s_{LS},a_{\tau})$} \\ \cmidrule{3-7}
    \multicolumn{1}{c}{$s_{LS}$}  & \multicolumn{1}{c}{$s_{SC}$} &  \multicolumn{1}{c}{$\theta_{off}$} &  \multicolumn{1}{c}{$\theta_{on}$} & $\tau_{neg}$ & $\tau_{neu}$ & $\tau_{pos}$\\ \midrule
    \multirow{3}{*}{$LS_{1}$} & \cellcolor{green!40} ${SC}_{L}$ & 0.75 & 1.0  & 0.10 & 0.50 & 0.40 \\
    & \cellcolor{yellow!50}${SC}_{M}$                           & 0.75 & 0.50 & 0.50 & 0.30 & 0.20 \\
    & \cellcolor{red!30}${SC}_{H}$                              & 0.50 & 0.25 & 0.60 & 0.30 & 0.20 \\ \midrule
    \multirow{3}{*}{$LS_{2}$} & \cellcolor{yellow!50}${SC}_{L}$ & 0.50 & 0.75 & 0.20 & 0.30 & 0.50 \\
    & \cellcolor{green!40} ${SC}_{M}$                           & 0.75 & 0.75 & 0.10 & 0.50 & 0.40 \\
    & \cellcolor{yellow!50}${SC}_{H}$                           & 0.75 & 0.50 & 0.50 & 0.30 & 0.20 \\ \midrule
    \multirow{3}{*}{$LS_{3,4}$} & \cellcolor{red!30} ${SC}_{L}$ & 0.25 & 0.50 & 0.10 & 0.30 & 0.60 \\
    & \cellcolor{yellow!50}${SC}_{M}$                           & 0.50 & 0.75 & 0.10 & 0.40 & 0.50 \\
    & \cellcolor{green!40} ${SC}_{H}$                           & 1.0  & 0.75 & 0.10 & 0.50 & 0.40 \\\bottomrule
   \end{tabular}
\end{table}

\begin{table}[h]
    \centering
    \caption{Reward $R_{1}(s,a)$ for calibrating self-confidence to learning stage. Green highlights represent cases of calibration, and yellow and red highlights represent cases of light and severe mis-calibration, respectively.}
    \begin{tabular}{cc|cccccc}
        \toprule
        \multicolumn{2}{c}{\textbf{TONE $a_{\tau}$:}} & \multicolumn{2}{c}{$\tau_{neg}$} & \multicolumn{2}{c}{$\tau_{neu}$} & \multicolumn{2}{c}{$\tau_{pos}$} \\
        \cmidrule(lr){3-4} \cmidrule(lr){5-6} \cmidrule(lr){7-8}
        \multicolumn{2}{c}{\textbf{ASSIST $a_{\theta}$:}} & $\theta_{off}$ & $\theta_{on}$ & $\theta_{off}$ & $\theta_{on}$ & $\theta_{off}$ & $\theta_{on}$ \\
        \midrule
        \multirow{3}{*}{${LS}_{1}$}
        & \cellcolor{green!40}${SC}_{L}$  & 0.43 & 0.65 & 0.78 & 1.00 & 0.70 & 0.91 \\
        & \cellcolor{yellow!50}${SC}_{M}$  & 0.78 & 0.57 & 0.61 & 0.39 & 0.52 & 0.30 \\
        & \cellcolor{red!30}${SC}_{H}$ & 0.65 & 0.43 & 0.39 & 0.17 & 0.22 & 0.010 \\
        \midrule
        \multirow{3}{*}{${LS}_{2}$}
        & \cellcolor{yellow!50}${SC}_{L}$  & 0.30 & 0.52 & 0.39 & 0.61 & 0.57 & 0.78 \\
        & \cellcolor{green!40}${SC}_{M}$  & 0.44 & 0.43 & 0.79 & 0.77 & 0.70 & 0.69 \\
        & \cellcolor{yellow!50}${SC}_{H}$ & 0.78 & 0.57 & 0.61 & 0.39 & 0.52 & 0.30 \\
        \midrule
        \multirow{3}{*}{${LS}_{3,4}$}
        & \cellcolor{red!30}${SC}_{L}$  & 0.010 & 0.22 & 0.17 & 0.39 & 0.43 & 0.65 \\
        & \cellcolor{yellow!50}${SC}_{M}$  & 0.22 & 0.43 & 0.48 & 0.70 & 0.57 & 0.78 \\
        & \cellcolor{green!40}${SC}_{H}$ & 0.65 & 0.43 & 1.00 & 0.78 & 0.91 & 0.70 \\
        \bottomrule
    \end{tabular}
    \label{tab:R_1}
\end{table}

\emph{Objective 2:} The reward $R_{2}(s,a)$ encourages worklload levels appropriate to the learner's current stage. This is consistent with learning stage theory from \cite{dreyfus_five-stage_2004}, which states that task-related cognitive demand declines as expertise grows. This trend is consistent with the learning-stage characterization reported in \cite{yuh_classification_2024}. Table~\ref{tab:R_2} implements this logic. At learning stages 3 and 4, low workload paired with no assistance earns the maximum reward of 1.0, reflecting that proficient learners should be able to land the drone manually with minimal cognitive demand. Conversely, at learning stage 1, low workload also earns 1.0, since a novice attempting a manual landing should experience an elevated workload, and the reward is highest when the workload is appropriately challenging for their current level. When the learner's workload is high, assistance is rewarded more than its absence, since the learner likely lacks the spare mental capacity to manage the task unaided. Only the assistance action enters $R_{2}$, since assistance is provided during the trial, whereas feedback is delivered only after the attempt.

\begin{table}[h]
    \centering
    \caption{Reward $R_{2}(s,a)$ for calibrating workload. Green highlighting indicates calibrated cases, while yellow and red highlighting indicate light and severe mis-calibration, respectively.}
    \begin{tabular}{cc|cc}
        \toprule
        \multicolumn{2}{c}{\textbf{ASSIST $a_{\theta}$:}} & {$\theta_{off}$} & {$\theta_{on}$} \\
        \midrule
        \multirow{3}{*}{${LS}_{1}$}
        & \cellcolor{red!30}${W}_{L}$  & 1.0  & 0.010 \\
        & \cellcolor{yellow!50}${W}_{M}$  & 0.60  & 0.40 \\
        & \cellcolor{green!40}${W}_{H}$ & 0.30  & 0.70 \\
        \midrule
        \multirow{3}{*}{${LS}_{2}$}
        & \cellcolor{yellow!50}${W}_{L}$  & 0.60  & 0.40 \\
        & \cellcolor{green!40}${W}_{M}$  & 0.45  & 0.55 \\
        & \cellcolor{yellow!50}${W}_{H}$ & 0.40  & 0.60 \\
        \midrule
        \multirow{3}{*}{${LS}_{3,4}$}
        & \cellcolor{green!40}${W}_{L}$  & 1.0  & 0.010 \\
        & \cellcolor{yellow!50}${W}_{M}$  & 0.70  & 0.30 \\
        & \cellcolor{red!30}${W}_{H}$ & 0.40  & 0.60 \\
        \bottomrule
    \end{tabular}
    \label{tab:R_2}
\end{table}

\emph{Objective 3:} The reward $R_{3}(s,a)$ incentivizes progression through the learning stages, as shown in Table~\ref{tab:R_3}. With automation off, reaching learning stage 3 or 4 earns the maximum reward of 1.0, with progressively lower rewards assigned for stages 1 and 2, encouraging the learner to advance toward expertise in manual mode. With automation on, the pattern inverts: learning stage 1 earns the maximum reward of 1.0, reflecting that a learner at this stage is most likely to benefit from assistance during the trial, while rewards decline across stages 2 through 4 as the need for assistance diminishes. As with Objective 2, only the assistance action enters $R_{3}$, since it directly shapes the learner's performance within each trial.

\begin{table}[h]
    \centering
    \caption{Reward $R_{3}(s,a)$ for advancing learning stage.}
    \begin{tabular}{c|cc}
        \toprule
        \textbf{ASSIST $a_{\theta}$:} & {$\theta_{off}$} & {$\theta_{on}$} \\
        \midrule
        ${LS}_{1}$   & 0.10  & 1.0 \\
        ${LS}_{2}$   & 0.55  & 0.45 \\
        ${LS}_{3,4}$ & 1.0  & 0.25 \\
        \bottomrule
    \end{tabular}
    \label{tab:R_3}
\end{table}

All reward function components are designed to fall within the range $[0,1]$.

\section{Cluster assignment and metrics of the change group}
\label{app:change_group}
Table~\ref{tab:changed_phase12_summary} summarizes the mean self-reported self-confidence, and mean self-reported workload over the final evaluation trials for change group participants in Table \ref{tab:changed_phase12_summary}.

\begin{table}[h]
\centering
\caption{Change group participants whose cluster assignment changed between phases 1 and 2. The provided information includes the assigned cluster for phase 1 $\alpha^{(1)}$ and assigned cluster for phase 2 $\alpha^{(2)}$ as well as the mean score, mean self-reported self-confidence, and mean self-reported workload over trials $k\in[33,37]$}
\label{tab:changed_phase12_summary}
\begin{tabular}{lccccc}
\toprule
Part. & $\alpha^{(1)}$ & $\alpha^{(2)}$ & Mean Score & Mean $SR_{SC}$ & Mean $SR_{W}$ \\
\midrule
1 & 1 & 2 & 937.6 & 72 & 80 \\
5 & 1 & 2 & 786.4 & 67 & 43 \\
7 & 2 & 1 & 872.0 & 89 & 6 \\
10 & 3 & 2 & 888.6 & 57 & 54 \\
12 & 3 & 2 & 574.0 & 71 & 83 \\
16 & 1 & 2 & 920.2 & 85 & 54 \\
18 & 1 & 2 & 941.8 & 81 & 34 \\
19 & 1 & 2 & 933.8& 89 & 44 \\
23 & 3 & 2 & 921.0 & 80 & 80 \\
24 & 3 & 1 & 733.6 & 49 & 20 \\
25 & 1 & 2 & 925.8 & 86 & 30 \\
28 & 1 & 2 & 801.4 & 73 & 80 \\
30 & 3 & 2 & 924.6 & 75 & 64 \\
32 & 3 & 2 & 780.0 & 42 & 79 \\
33 & 3 & 1 & 294.8 & 10 & 90 \\
39 & 1 & 2 & 794.6 & 87 & 35 \\
\bottomrule\end{tabular}
\end{table}

\vfill

\end{document}